\documentclass[12pt,a4paper]{article}
\usepackage[english]{babel}
\usepackage{epsfig}

\begin{document}

\newcommand{\ptop}{{p_{\top}}}
\newcommand{\cz}{{H^{\pm}}}
\newcommand{\wb}{{W^{\pm}}}
\newcommand{\wbp}{W^+}

\thispagestyle{empty}

  \newcommand{\ccaption}[2]{
    \begin{center}
    \parbox{0.85\textwidth}{
      \caption[#1]{\small{{#2}}}
      }
    \end{center}
    }

\vspace{0.5cm}

\begin{center}
{\Large { \sf
{\bf Charged Higgs Contribution into $t \bar b$-pair Production
in Hadronic Collisions }}}
\end{center}

\vspace{1cm}

\begin{center}
M.V.~Foursa$^1$, D.A.~Murashev$^2$ and S.R.~Slabospitsky$^3$
\end{center}

\vspace{1cm}

\begin{center}
{\it
State Research Center \\
Institute for High Energy Physics, \\
Protvino, Moscow Region 142284 \\
RUSSIA }

\end{center}

\vspace*{1.5cm}

\begin{abstract}
We investigate the charged Higgs boson contribution into $t\bar b$-pair
production in $pp$-collisions at LHC. It is shown that due to the $\cz$-boson
exchange the total yield of $t\bar b$ is modified significantly for small and
large values of $\tan\beta$. However, for the small values of $\tan\beta$
one should expect
also  the production of right-handed top quark contrary
to pure left-handed $t$-quark production through $\wb$-boson exchange only.
This fact provides a possibility to separate $\cz$ and $\wb$
contributions by means of the investigations of angular distributions
of top decay products. The detailed simulation of the signal and
relevant background processes is performed.
\end{abstract}

\vspace*{1cm}
\vfill

\begin{center}
 Protvino,~~~2000
\end{center}

\rule{3cm}{0.5pt}

\noindent
$^1$~fursa$@$mx.ihep.su \\
$^2$~murashev$@$mx.ihep.su \\
$^3$~slabospitsky$@$mx.ihep.su \\

\newpage

\section{Introduction }
\vspace{-2mm}
The existence of the charged Higgs boson~$\cz$ is predicted by many extensions
of the Standard Model (see, for example,~\cite{higrev}).
The search for a charged Higgs boson is carried out in $e^+ e^-$ annihilation
at LEP-2 collider at CERN~\cite{lep-exp} as well as at the Tevatron in the top
quark decays~\cite{tev-exp}. The search for the $\cz$-boson will be one of
the main experimental tasks at future LHC
machine~\cite{Beneke:2000hk,Djouadi:2000gu}.
The main channels for $\cz$-boson search are the reactions
 of $gb\to tH^-$ and $gg\to t\bar bH^-$b~\cite{Djouadi:2000gu, higprod, cms1}.

In the present paper we consider an additional possibility to study
a signal from charged Higgs boson in the quark-antiquark annihilation
subprocess:
\[
 q\bar q'\to H^+\to t\bar b \; .
\]
Namely, we consider the charged Higgs boson contribution to the process of
$t \bar b$-pair production in proton-proton collisions
\begin{eqnarray}
 pp \, \to \, t \, \bar  b \, X.  \label{mainproc}
\end{eqnarray}

Note, that $t\overline b$ quark production through  $W$-boson
exchange in the $s$-channel had been considered earlier
(see~\cite{Beneke:2000hk,wstar, Heinson, tait}).
This reaction plays a significant role in the study
 of electroweak $t$-quark production. In particular,
this process  is very important for the investigations of
electroweak vertex of $tWb$ interactions~\cite{Beneke:2000hk}.

The New Physics beyond the Standard Model (SM) can modify the nature of
$t$-quarks interactions (see~\cite{Beneke:2000hk} and references therein).
In particular, the contribution of charged Higgs boson into 
process~(\ref{mainproc}) can be considered as a manifestation of
New Physics. The existence of the $\cz$-boson leads not only to the 
modification of the total cross section for $t \bar b$-pair production, but 
also to the change of angular distributions of top quark decay products.
The detailed consideration of the contribution due to several forms
of the New Physics into $t \bar b$-pair hadronic production as well
as the analysis of the top quark polarization properties resulted
from such new interactions can be found, for example, in~\cite{tait}.

Here we investigate the charged Higgs boson production in
proton-proton collisions at future CERN~LHC machine at the
energy of $\sqrt s=14$~TeV. Certainly, the strategy for the search
for the $\cz$-boson
production is determined, in particular, by the Higgs boson mass
(see~\cite{Djouadi:2000gu}). For relatively
light $\cz$, say, $m_{\cz} < m_t$, the most promising possibility is the
investigation of $t$-quark decay into $\cz$ and $b$-quark. In the present
paper we assume that the charged Higgs is heavier than the top quark.
As a result, we consider the top quark decays into $W^{\pm} b$ pair.
Moreover, we consider only leptonic decays of $W^{\pm}$-boson, because for
hadronic $\wb$ decays it is
very difficult to separate a signal from a huge QCD background.

 Note, that typical
differential distributions ($p_{\top}$ and $\eta$) of the $t$-quark and its
decay products are practically the same as for the SM production of
$t \bar b$-pair (via $\wbp$-boson exchange only). Therefore, for the
separation of the $\cz$ contribution we explore $t$-quark polarization
properties, which are different for $W^+ \to t \bar b$ and $H^+ \to t \bar b$
contributions for small values of $\tan \beta$.
We find the specific kinematic cuts, which provide the separation of $\cz$ and
$W^{\pm}$ contributions in reaction~(\ref{mainproc}).

The paper is organized as follows.
In Section~2 we present an explicit form of matrix elements
squared for the considered process.
The behavior of the total cross section production of
$t \bar b$-pair as a function of $m_H$ and $\tan\beta$ is considered
in Section~3. The differential distributions on $\ptop$ and $\eta$
as well as the angular distributions of top decay products are also discussed
in this Section. A detailed simulation of the signal and
relevant background processes is given in Section~4.
The main results are summarized in the Conclusion.

\vspace{-2mm}
\section {Matrix elements calculations }
\vspace{-2mm}
The Lagrangian describing the $\bar t \cz b$-vertex in the two doublet Higgs model
has the following form~\cite{higrev, Djouadi:2000gu, Beneke:2000hk}:
\begin{eqnarray}\label{lagrang}
{\cal L}=\sqrt{\frac{G_F}{\sqrt{2}}}H^+
 \{\tan\beta\overline U_L V_{ij} M_D D_R + \cot\beta\overline
U_R M_U V_{ij} D_L + \tan\beta\overline N_LM_LL_R\},
\end{eqnarray}
where the symbols $U$ and $D$ refer to 'up' and 'down'-type quarks,
while $N$ and $L$ correspond to the charged
lepton and neutrino, $V_{ij}$ is the Cabbibo-Kobayashi-Maskawa matrix element,
$M_D$ and $M_U$ are the quark masses, and $M_L$ is the mass of
charged leptons, $\tan\beta$ is the ratio of values of vacuum
expectation value for two Higgs doublets, $G_F$ is the Fermi constant.

The subprocesses of $t \bar b$-pair production through $\wb$ and
$\cz$ exchange,
\begin{equation}\label{star1}
q_1\overline q_2\to(H^++W^+)\to t\overline b,
\end{equation}
is described by two diagrams presented in Fig.~1.

The corresponding matrix element squared is presented  as a sum of three
 terms, corresponding
to $\cz$-boson ($M_H$) and $\wbp$-boson ($M_W$) exchanges and their
interference ($M_I$):
\begin{eqnarray}\label{m22}
|M_{2 \to 2}|^2 = |M_H|^2 \, + \,  |M_W|^2 \, + \,  |M_I|^2,
\end{eqnarray}
where
\begin{eqnarray*}
|M_H|^2&=&\frac{16G_F^2|V_{12}|^2|V_{tb}|^2}
{\left(\hat s-m_H^2\right)^2+\Gamma_H^2m_H^2}
\left[\left(m_t^2\cot^2\beta+m_b^2\tan^2\beta\right)
(p_t p_b)-2m_b^2m_t^2\right]\\
&&\times\left[\left(m_1^2\cot^2\beta+m_2^2\tan^2\beta\right)
(p_1 p_2)-2m_1^2m_2^2\right], \\
|M_W|^2&=&\frac{128m_W^4G_F^2|V_{12}|^2|V_{tb}|^2}
{\left(\hat s-m_W^2\right)^2+\Gamma_W^2m_W^2}(p_t p_2)(p_b p_1), \\
|M_I|^2&=&\frac{32G_F^2|V_{12}|^2|V_{tb}|^2m_tm_bA}{A^2+B^2} \\
&& \times [-m_1^2\cot^2\beta(p_t p_2)
 +m_2^2(p_t p_1)+m_1^2(p_b p_2)-m_2^2\tan^2\beta(p_b p_1)],
\end{eqnarray*}
where $\hat s$ is the total energy squared of colliding partons,
$V_{ij}$ is the Cabbibo-Kobayashi-Maskawa matrix element,
$\Gamma_H$ is the $H^+$-boson decay width,
$m_i$ and $p_i$ are the quark masses and 4-momenta, respectively,
$m_W$ and $\Gamma_W$ are the W-boson mass and decay width,
$A=(\hat s-m_W^2)(\hat s-m_H^2)+\Gamma_Wm_W\Gamma_Hm_H$,
$B=(\hat s-m_W^2)\Gamma_Hm_H-(\hat s-m_H^2)\Gamma_Wm_W$.

Bearing in mind the study of the $t$-quark polarization properties in
reaction~(\ref{mainproc}), we also perform the calculations of subprocesses
(\ref{star1}) with the consideration of further top quark decay.
Namely, we calculate the matrix element squared for the
subprocess $2\to4$:
\begin{equation}\label{star2}
q_1\overline q_2\to(H^++W^+)\to \overline bt(\to bW^+)
\longrightarrow b\overline bl^+\nu_l.
\end{equation}
Similarly to subprocesses (\ref{star1}), we present the $|M_{2\to4}|^2$
as a sum of three terms corresponding to the $\cz$ and the $\wb$
exchanges and their interference:
\begin{equation}
 |M_{2\to4}|^2 = |M_{H}|^2 \, + \, |M_{W}|^2 \, + \, |M_{I}|^2, \label{m24}
\end{equation}
where
\begin{eqnarray*}
|M_{H}|^2&=&\frac{2048m_W^4G_F^4|V_{tb}|^4|V_{12}|^2(p_b k_1)}
{((\hat s-m_H^2)^2+\Gamma_H^2m_H^2) \, C_W  C_t } \\
&&\nonumber\times
\{(q_1,q_2)[m_1^2\cot^2\beta+m_2^2\tan^2\beta]-2m_1^2m_2^2\}\\
&&\nonumber\times\{m_b^2\tan^2\beta[2(k_2 p_t)(p_{\bar{b}} p_t)
-(k_2 p_{\bar{b}})p_t^2] \\
&& +m_t^4\cot^2\beta(k_2 p_{\bar{b}})-2m_b^2m_t^2(k_2 p_t)\}, \\
|M_{W}|^2&=&\frac{8192m_W^8G_F^4|V_{tb}|^4|V_{12}|^2(p_b k_1)
(p_{\bar{b}},q_1)}
{((\hat s-m_W^2)^2+\Gamma_W^2m_W^2) \, C_W  C_t } \\
&&\times [2(q_2 p_t)(k_2,p_t)-(q_2 k_2)p_t^2], \\
|M_{I}|^2&=&\frac{4096m_W^6G_F^4|V_{tb}|^4|V_{12}|^2Am_1m_2}
{(A^2+B^2)  \, C_W C_t } \\
&&\times
\{m_t^2\cot^2\beta[(q_1,p_{\bar{b}})(p_t k_2)+(q_1 p_t)(p_{\bar{b}} k_2)
-(q_1 k_2)(p_{\bar{b}} p_t)]\\ &&\nonumber
+m_t^2[(q_2 k_2)(p_{\bar{b}} p_t)-(q_2 p_{\bar{b}})(p_t k_2)
-(q_2,p_t)(p_{\bar{b}},k_2)]\\ &&\nonumber
+m_b^2\tan^2\beta[2(q_2 p_t)(p_t k_2)-(q_2 k_2)p_t^2]\\
&&\nonumber+m_b^2[(q_1 k_1)p_t^2-2(q_1 p_t)(p_t k_2)]\},
\end{eqnarray*}
where $A$ and $B$ is defined in~(\ref{m22});
$C_W = (p_W^2-m_W^2)^2+\Gamma_W^2m_W^2$ and
$C_t = (p_t^2-m_t^2)^2+\Gamma_t^2m_t^2$, $p_W$ and $p_t$ are the momenta
of top quark and $W$-boson, respectively; $k_1$ and $k_2$ are the momenta
of neutrino and charged lepton.

Note, that the matrix element squared~(\ref{m24}) corresponding to
subprocess~(\ref{star2}) is calculated for the first time.

\vspace{-2mm}
\section {Total cross sections and differential distributions}
\vspace{-2mm}
In our calculations we use the parton distributions from~\cite{Lai:2000wy}.
The evolution parameter $Q^2$ is chosen to be $Q^2=\hat{s}$,
where $\hat{s}$ is the total energy squared of colliding partons. In
our calculations we use the following values of the $b$ and $t$-quark masses
\begin{equation}
 m_b=4.5 \,\,{\rm GeV}\,\,\cite{rpp} \quad {\rm and} \quad m_t = 173.8
\,\,{\rm GeV}.   \label{mbt}
\end{equation}

For the fixed value of the charged Higgs boson mass the largest cross section
is expected at large and small values of $\tan\beta$.
The behavior of the cross section for reaction~(\ref{mainproc})
as a function of the
$H^{\pm}$-boson mass and $\tan\beta$ is presented in
Fig.~2. Note, that the cross section
reaches its maximum at $m_H\sim 200$~GeV due to the pole
character of cross section under consideration (see~(\ref{m22})).

The $\cz$-boson contribution into reaction~(\ref{mainproc})
leads to the increasing of the $t \bar b$-pair cross
section production. However, the modification of the ``observable'' number of
events, $N_{ev}(W b \bar b) \sim \sigma(t \bar b) \times
{\rm B}(t \to b W^+)$, depend on
$m_{\cz}$. For the $\cz$-boson lighter than top quark ($m_{\cz} < m_t$), the
branching fraction of $t \to b W$ decay (B$(t \to b W)$) may
significantly decrease~\cite{Djouadi:2000gu, Beneke:2000hk}.
As a result, we should expect the decrease of $N_{ev}$ as compared to the
SM case. For heavy $\cz$-boson ($m_{\cz} > m_t$) there is no $t \to b H^+$
decay channel. Therefore, one has B$(t \to b W^+) \approx 1$ and
 we should expect the increase of $W^+ \, b \, \bar b$ yield.

The ``experimentally seen'' cross sections for
$t$-quark production in the reaction~(\ref{mainproc}), i.e.
\begin{equation}\label{star3}
\sigma(pp\to \bar bt(\to bW)X)=
\sigma(pp\to t\bar bX)\times{\rm B}(t\to bW)
\end{equation}
are presented in Fig.~3 for two values of $m_H$=90 and 200 GeV.
Note, that for the small and large values of $\tan\beta$ the $\cz$-boson
contribution leads to a noticeable modification of the $t \bar b$-pair
production, while for an intermediate range of $\tan\beta=0.8\div 20$
the $\cz$ contribution becomes negligible.
The differential
distributions of the final particles ($t, W, b, l$) on
transverse momentum and pseudorapidity calculated at the parton level
are shown in Figs.~4,~5. As is shown in the figures, the differential
distributions for the cases of $H^+$ and $W^+$ exchange
have practically the same shapes.
Therefore, the contribution of New Physics may lead only to the deviation
of the expected number of events. Therefore, to elucidate the nature of
possible deviation from the SM predictions, one needs to examine additional
quantities which have a differential behavior for the SM and beyond SM cases.

For this purpose we explore the polarization properties of the $t$-quark,
widely  considered in the literature (see, for example,
~\cite{Beneke:2000hk, Jezabek:1989ja, Mahlon:1997pn} and references
therein).
It is well known that subprocess~(\ref{star1}) leads to the
production of almost left-handed $t$-quark~\cite{Beneke:2000hk,tait}.
At the same time the charged Higgs boson contribution
leads to the production of right(left)-handed $t$-quark for small (large)
values of $\tan\beta$ (see~\cite{Djouadi:2000gu}).
Note, that for the left-handed $t$-quark the $b$-quark (the
charged lepton) should fly mostly alongside (opposite) to the direction
of the top quark momentum (see~\cite{Beneke:2000hk,tait}).
Naturally, for the right-handed quark one has the reverse situation.

Hereafter, in order to separate the $\wb$ and $\cz$ contributions,
we examine the angular distributions,
\begin{equation}\label{cs}
\frac{dN}{dcos\theta^*},
\end{equation}
where $\theta^*$ is the angle in the top rest frame between the direction
of the top quark momentum and the momentum of the final particle from the
top quark decay~$t \to b l \nu$.

In Fig.~6 we
present the corresponding angular distributions of the $b$-quark and charged
lepton from $t$-quark decay calculated by means of the matrix
element (\ref{m24}) separately for only $\wbp$ or $\cz$ exchange
(see the histograms ``a, b, e, f'' in the Fig.~6). For evaluating
these distributions, we set $m_{\cz} = 200$~GeV and $\tan \beta = 0.1$.

\vspace{-2mm}
\section { Signal and background calculations }
\vspace{-2mm}
We perform the detailed simulation of process (\ref{mainproc})
with the subsequent top quark decays into electron and muon
($t \to b e \nu_e$ and $t \to b \mu \nu_{\mu}$) and relevant
 background reactions for the three-year low luminosity run of LHC
\begin{equation}
 \sqrt{s} = 14 \,\,{\rm TeV} \quad {\rm and} \quad \int {\cal L} dt = 30\,\,
{\rm fb}^{-1}. \label{lum}
\end{equation}
As an example we choose the values of the charged Higgs boson parameters
as follows:
\begin{equation}
 m_{\cz} \, = \, 200\,\,\,{\rm GeV} \quad {\rm and} \quad
\tan\beta \, = \, 0.1. \label{mtb}
\end{equation}
Note, that our result is not very sensitive to the $m_H$ value, however, the
chosen small value of $\tan\beta$, namely, $\tan\beta<0.2$, is important for
further analysis.

For the simulation of the signal and background processes we use the
{\bf TOPREX} {\bf 2.51} event generator~\cite{toprex}. At present  the TOPREX
provides the generation at the parton level, the production of $t \bar t$-pair,
three processes of single top (``$Wg$'', ``$t W$'', and ``$W^*$'')
and $W b \bar b$  production with subsequent top quark
decay into $bW$ and $W$ into fermion anti-fermion pair.
One can generate also several processes of top quark production via anomalous
$t$-quark interactions~\cite{Gouz:1999rk}.

We use the PYTHIA~6.134~\cite{PYTHIA} for simulation of quarks and
gluons hadronization. We perform the simulation with taking into account
the ability of CMS detector at LHC~\cite{cms}.
 For ``fast'' detector simulation all the events are passed
through the package CMSJET~4.703~\cite{CMSJET}.
As a result, the final event contains the information about momenta of
``detected'' photons,
charged leptons ($e,\mu$), hadronic jets and missing energy
$E_{\top mis}$ (see~\cite{CMSJET} for details).
The efficiencies for $b$-tagging of jets originating
from $b$-quark, $c$-quark and light partons ($u, d, s$-quarks and gluon) are
about 60\%, 10\%, and $1\div2 \%$, respectively~\cite{CMSJET}.
In what follows we refer to the $b$-tagged jet as the $B$-jet.

For our choice of the $\cz$-boson parameters, (see~(\ref{mtb})),
the cross section for ($t \bar b \, + \, \bar t b$)  production in
reaction~(\ref{mainproc}) due to $\cz$-boson contribution
at $\sqrt{s} = 14$~TeV is equal to
\begin{equation}
 \sigma(\cz) \, = \, 9.7\,\,\,{\rm pb}.
\end{equation}

There are several sources of the background to the considered
process~ (\ref{mainproc}).  We present here the cross section values of these
processes evaluated at LO approximation.
The consideration of the higher order corrections
is given, for example, in~\cite{Beneke:2000hk}. These background processes
are as follows (all the cross section values imply the summation on particles and
anti-particles):
\begin{itemize}
\item[$\bullet$] three processes of single top production
\begin{eqnarray*}
 q \bar q' \to W^\pm \to t\bar b, && \sigma(W^*) =   7.5 \,\,{\rm pb}, \\
 g q \to q't \bar b,            && \sigma(Wg)  = 180   \,\,{\rm pb}, \\
 g b \to t W,                   && \sigma(tW)  =  60   \,\,{\rm pb}
\end{eqnarray*}
\item[$\bullet$] $t \bar t$-pair production:
\hspace{0.5cm} $ gg(q\bar q) \to t\bar t, \quad \sigma(t \bar t) = 600$~pb
\item[$\bullet$] $W \, b \,  \bar b$ production: \hspace{0.5cm}
$ q \bar q' \to W \, b \, \bar b, \quad \sigma(W b \bar b) = 360$~pb
\item[$\bullet$] $W+jets$ production (generated by PYTHIA)
\begin{eqnarray*}
 \sigma(W+jets) = 59000\,\,{\rm pb}
\end{eqnarray*}
\end{itemize}
\vspace{-3mm}
Signal process~(\ref{mainproc}) and all the background reactions (except
 $W$+jets) are simulated by using TOPREX generator.
For evaluation of the last process ($W$+jets) the PYTHIA parameter
$k_{\top min}$ is chosen to be equal to 2 GeV (i.e. CKIN(1)=2~\cite{PYTHIA}).

Process (\ref{mainproc}) of $t \bar b$ pair production with the subsequent
top decay into $b l \nu$ is characterized by the presence in the final state
of one charged isolated lepton (from $W$-boson decay),
the missing energy (neutrino) and two hard $B$-jets from $b$-quarks.

The appropriate cuts providing a signal selection from the background
processes, are considered in detail in ~\cite{Beneke:2000hk}.
In particular, these cuts include the requirement of two ``hard''
$B$-jets:
\begin{equation}
{\rm ``hard''-cut}: \,\, p_{\top}(B_1, B_2) \, \geq p_{T0} \, \sim \, 75 \,\,
{\rm GeV} \label{hard}
\end{equation}
However, this cut leads to an essential modification of the form of the corresponding
$\cos \vartheta^*$ distributions (\ref{cs}) of the top quark decay products.

Indeed, in Fig.~6 we present the angular distributions of the
$b$-quark (from $t$-quark decay) evaluated at the parton level before and
after the cut on the $b$-quark transverse momentum
(see the histograms ``b, c, f, g'' in Fig.~6).
 One can see that after application
of the cut (\ref{hard}) the form of the angular distribution of $b$-quark,
originating
from the decay of right-handed $t$-quark (produced via $\cz$ exchange)
changes dramatically and even becomes qualitatively similar to that of
$b$-quark from left-handed top decay (produced via $\wbp$ exchange).
In view of actual possibilities of the detector it will be even
more difficult to distinguish these two distributions.

Therefore, we propose another $p_{\top}$-cut (``soft-hard'') on the final
$B$-jets. Namely, we require that $\ptop$ of one $B_1$-jet (from top quark)
should not exceed some value of $p_{T1} = 100$~GeV, while $\ptop$ of the
other $B_2$-jet
should be greater than $p_{T0} = 75$~GeV:
\begin{equation}\label{newcut}
{\rm ``soft-hard''-cut}: \,\, p_{\top}(B_1\,\,{\rm from \,\, t}) \,
\leq p_{T1} \,\,\,\,{\rm and} \,\,\, p_{\top}(B_2) \geq p_{T0}.
\end{equation}
It is seen from the histograms ``d'' and ``h'' in Fig.~6 that this
``soft-hard'' cut~(\ref{newcut}) saves the forms of angular distributions
of the decay products of $t$-quarks, produced through
 $\cz$ and $\wb$ exchange.

In the further analysis  we apply both variants~(\ref{hard}) and~(\ref{newcut})
of $\ptop$ cuts.
Thus, for signal/background separation we require
\begin{itemize}
\item[1)] one and only one
isolated lepton (with $p_\top>10$ GeV) and at least two hadronic jets
(with $p_\top>20$ GeV and pseudorapidity $|\eta|<4.5$),
\item[2)]  explicitly two b-tagged jets
(with $p_\top>25$ GeV and $|\eta|<2.5$) and no other hadronic jets,
\item[3)] the transverse momentum of the reconstructed $W$-boson and two
$B_1$ and $B_2$ jets should not exceed of
10~GeV, $|\vec p_T(W B_1 B_2)| \leq 10$~GeV,
\item[4)] $H$(``hard'')-cut on $\ptop$ of $B$-jets ($\ptop(B_1, B_2) \geq75$ GeV),
\item[4')]$SH$(``soft-hard'')-cut on $\ptop$ of $B$-jets (one $B_1$-jet with
$\ptop \leq 100$~GeV and second $B_2$-jet with $\ptop \geq 75$~GeV),
\item[5)] the reconstructed mass of the $t$-quark should be
within $150\div200$ GeV, \\ $|M_{rec}(BW)-m_t|<25$ GeV.
\end{itemize}

As usual, using the four-momentum of charged lepton and the transverse momentum
of missing energy for the reconstruction of $W$-boson momentum, we get two
solutions for $p_{rec}(W)$. For further reconstruction of top quark
four-momentum, $p_{rec}(t)$,  we should examine both two $B$-jets for the
case of $H$-cut~(\ref{hard}). As a result, we obtain four combinations for
the reconstructed momentum of the $t$-quark.
For $SH$-cut~(\ref{newcut}) we assume that the $B_1$-jet with the smallest
transverse momentum results from the $t$-quark decay and we have
two solutions for $p_{rec}(t)$. Then, the invariant mass of
the ($W \, B$)-system, $M(WB)$, with the value nearest to
$m_t = 173.8$~GeV is treated as a reconstructed value of the
$t$-quark mass. The corresponding $p_{rec}(t)$ is considered
as a reconstructed $t$-quark momentum.

The resulted efficiencies after application of all the cuts to the signal
and the background  are given in Table~1. One can see that new
$SH$~cut~(\ref{newcut}) provides slightly better efficiency
for the signal reconstruction. However, the background
suppression becomes also worth as compared to the application of old
$H$-cut~(\ref{hard}). The resulted number of reconstructed events (for
$\int {\cal L}dt = 30$~fb$^{-1}$) and the corresponding signal-to-background
ratios are given in Table~2. It is seen that the application of the old and new
 $\ptop$-cuts result in almost the same $S/B$ ratios.
Therefore, both of the two variants of $\ptop$-cut ($SH$ and $H$) provide a rather
well reconstruction of the top quark in reaction~(\ref{mainproc}).

The distributions on the
reconstructed top quark mass are presented in Fig.~7. The symbol ``SM''
corresponds to calculations with the SM case, while the symbol ``SM+$\cz$''
corresponds to the SM and $\cz$-boson contribution.
The standard fit gives
the following values of the reconstructed $t$-quark mass~(in GeV):
\begin{center}
\begin{tabular}{|l|l|l|} \hline
              & SM & SM + $H^{\pm}$         \\ \hline
 S-H & $172.4 \pm 11.8$ & $171.6 \pm 11.4$  \\ \hline
 H   & $172.8 \pm 11.2$ & $172.4 \pm 10.8$  \\ \hline
\end{tabular}
\end{center}

Now we proceed to the separation of the $\cz$-boson contribution into
 reaction~(\ref{mainproc}). For this purpose we explore the difference
in $\cos \vartheta^*_l$-distributions resulted from different polarizations
of the $t$-quark, produced within the SM (only $W$ exchange) and within
$\wb + \cz$ exchange (see Fig.~6). The distributions on
$\cos \vartheta^*_l$, calculated for the sum of the signal and
background events
are given in Fig.~9. The upper two histograms correspond to new
$SH$-cut, while two lower histograms are obtained by application
of old $H$-cut.
We fit these distributions by linear dependence on
$\cos \vartheta^*_l$:
\begin{equation}\label{fit}
\frac{d \, N}{d \, \cos \theta^*_l} \propto 1 +   \alpha \cdot \cos \theta^*_l
\end{equation}
The results of the fit are given in Table~\ref{tab3}.
It is evident that the new $SH$-cut~(\ref{newcut}) is more sensitive
 to $\cz$-boson contribution. Indeed, the presence of charged
Higgs leads to change of the sign of the slope of
$\cos \vartheta^*_l$-distribution, while
the application of the old $H$-cut leads only to modification
of the slope~$\alpha$ (see Table~\ref{tab3}).

Certainly, this result depends on the relative value of the $\cz$-boson
contribution. Indeed, for a larger value of the $\tan\beta$
we should expect the decreasing of charged Higgs contribution and, as a result,
the values of $\alpha$ should be more close to SM expectations. Note, that
for the large value of $\tan\beta > 10$, where we have a noticeable
charged Higgs contribution, the produced top quark should be left-handed.
 As a result, the angular distribution should be the same as in the SM case.

\vspace{-2mm}
\section{Conclusion}
\vspace{-2mm}
In the present paper the contribution from charged Higgs bosons into the
process of electroweak production of $t \bar b$-pair at LHC is considered
and analyzed in detail. The expressions for matrix elements squared for
the corresponding subprocesses
are obtained and the role of the $t$-quark polarization is investigated.
The cross sections and angular distributions at a parton level are calculated.
The simulation of the signal and relevant background processes by means
of PYTHIA in view of opportunities of the CMS detector at LHC is also
performed.

We show that the differential distributions ($p_{\top}$ and $\eta$) of
the $t$-quark and its decay products are practically the same as for
the SM production of $t \bar b$-pair. As a result, by using a conventional
way for the top quark separation from the background, it would be difficult
to distinguish the $\cz$-boson from $\wb$-boson exchange.
 At the same time the produced top quark
through charged Higgs exchange has different polarization in comparison
to the SM case. At small values of $\tan\beta$, one should expect
the production of the right-handed $t$-quarks.
The corresponding angular distributions of leptons essentially differ
from those predicted by the SM.

In order to separate the $\wb$ and $\cz$ contributions into
reaction~(\ref{mainproc}) we propose the new $\ptop$-cut~(\ref{newcut}) for
$b$-tagged hadronic jets. This ``soft-hard'' $\ptop$-cut provides
 not only the $t$-quark signal discrimination from the background
 processes, but also the selection of  the contribution to the relevant
process from a charged Higgs boson.

Certainly, the proposed $\ptop$-cut can help only for the small values of
$\tan\beta \,\, < 0.2$. For the larger values
(i.e., for $\tan\beta > 0.2$) one
needs find other variables, which should be sensitive to charge Higgs boson
 contribution into reaction~(\ref{mainproc}).
In principle, we can explore the SM prediction for the
difference in top and anti-top quark production
 cross sections in reaction~(\ref{mainproc}). This difference results
 due to the absence of the valence anti-quark contributions in the
proton-proton collisions~\cite{Heinson, Jikia:1992yd}.
At the same time, the main contribution to the $\cz \to t \bar b$ process
comes from the interaction of charmed and strange quarks from initial
hadrons. Since in the $pp$-collisions we have equal numbers of the $c$, $s$
quarks and anti-quarks, we should expect $\sigma(H^+ \to t \bar b)
\approx \sigma(H^- \to \bar t b)$. Thus, the study of asymmetry in $t$-
and $\bar t$-quarks production in reaction~(\ref{mainproc})
may provide an additional  possibility to separate the charged Higgs boson
contribution.

\vspace{4mm}
\noindent {\bf Acknowledgment}\\

\noindent We are thankful to E.~Boos, D.~Denegri, V.~Drollinger, V.~Ilyin,
A.~Kostritsky, N.~Krasnikov,
M.~Mangano, A.~Nikitenko, V.~Obraztsov, L.~Sonnenschein, N.~Stepanov and
T.~Tait for fruitful discussions. The work of S.R.S. was supported, in part,
by Russian Foundation for Basic Research,
projects no.~99-02-16558.

\vspace{1cm}

\newpage


\begin{table}[ht]
\begin{center}
\caption{The efficiencies (in \%) of the signal ($\cz$) and background 
separation after application of cuts. The symbols $SH$ and $H$ correspond
to the usage of new~(\ref{newcut}) and old~(\ref{hard}) $\ptop$-cut. }
\vspace{0.4cm}
\begin{tabular}{|l|rrrrrrl|} \hline
 &  $H^{\pm}$ &  $W^{\pm}$ & $Wg$ & $Wt$ & $t \bar t$ & $W b \bar b$
      & $W+$jets  \\ \hline
  SH & 0.77 & 0.5 & 0.012 & 0.008 & 0.003 & 0.02 & $5\cdot10^{-5} $
   \\ \hline
  H   & 0.26 & 0.4 & 0.007 & 0.003 & 0.003 & 0.01  & --    \\ \hline
\end{tabular}
\label{tab1}
\end{center}
\end{table}

\begin{table}[ht]
\begin{center}
\caption{The number of events resulted after application of all the cuts.
In the calculations the total integrated luminosity of
$\int {\cal L}dt = 30$~fb$^{-1}$ is assumed. The symbols $W/B$ and
$H/B$ correspond to signal-to-background ratios, calculated within the SM
framework as well as for the SM and $\cz$-boson contribution, respectively.
The symbol SH(H) stands for the usage of new (old) $\ptop$-cut. }
\vspace{0.4cm}
\begin{tabular}{|l|rrrrrr|} \hline
     & $H^{\pm}$ & $W^{\pm}$ & SM & SM + $H^{\pm}$ &
 $\frac{W}{B}$ & $\frac{H}{B}$ \\ \hline
 SH       & 470 & 260 & 960 & 1430 & 0.37 &  0.49 \\ \hline
  H         & 260 & 220 & 610 &  870 & 0.55 & 0.43 \\ \hline
\end{tabular}
\label{tab2}
\end{center}
\end{table}

\begin{table}[ht]
\begin{center}
\caption{The result of the fit of $\frac{d \, N}{d \, \cos \theta^*_l}$
distribution by the function of $( 1 +   \alpha \cdot \cos \theta^*_l)$.
The symbol ``SM'' corresponds to the calculations with the SM case,
 while the symbol ``SM+$\cz$''
corresponds to the SM and $\cz$-boson contribution. }
\vspace{0.4cm}
\begin{tabular}{|l|l|l|} \hline
              & $\alpha$(SM) & $\alpha$(SM + $H^{\pm}$) \\ \hline
  SH & $-0.29 \pm 0.06$ & $+0.21 \pm 0.05$ \\ \hline
  H   & $-0.98 \pm 0.05$ & $-0.46 \pm 0.06$ \\ \hline
\end{tabular}
\label{tab3}
\end{center}
\end{table}

\newpage
\begin{figure}[ht]
\begin{center}
\epsfig{file=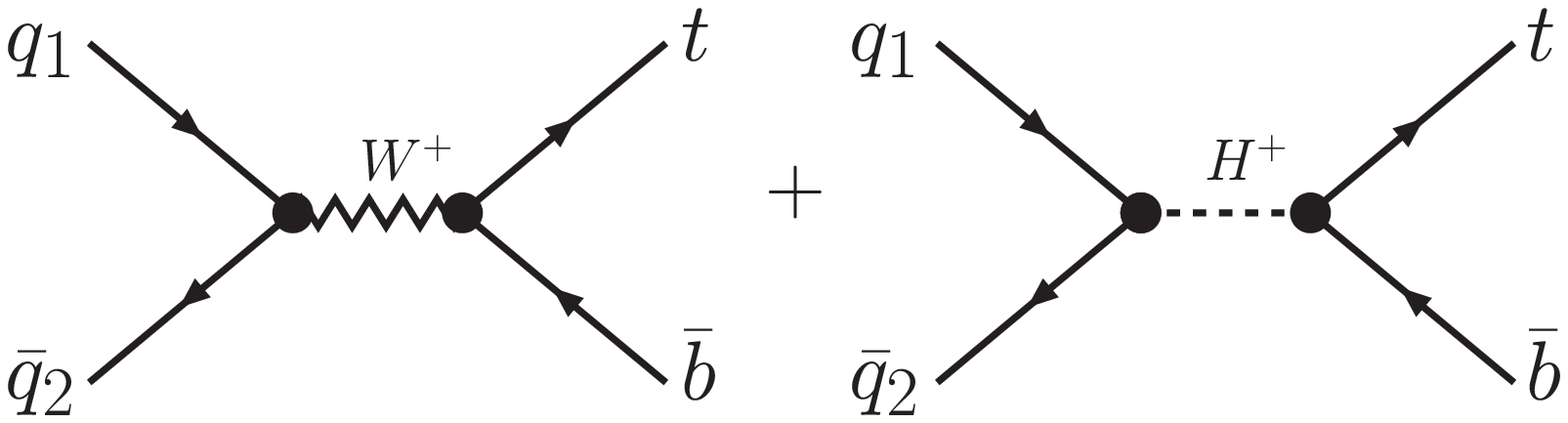,width=8cm,clip=}
\ccaption{}{ Feynman diagrams for the subprocess
$q_1\overline q_2 \, \to \, (W^{\pm} \cz) \, t\overline b$}
\end{center}
\end{figure}

\begin{figure}[ht]
\begin{center}
\epsfig{file=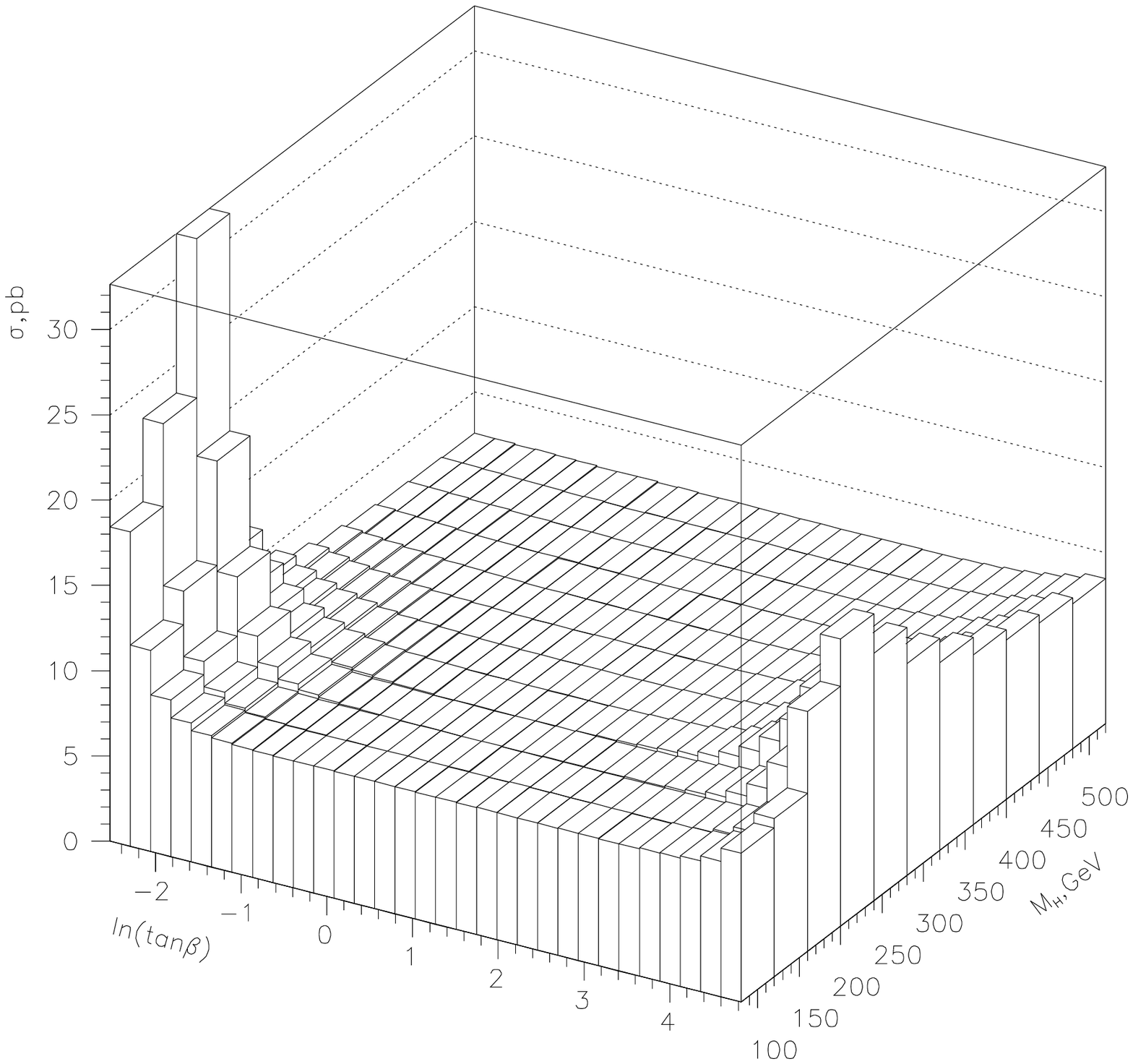,width=10cm,clip=}
\ccaption{}{ The behavior of the total cross section for
 ($t \bar b \, + \, \bar t b$)-pair 
production in $pp$-collision at $\sqrt{s} = 14$~TeV as a function of 
$\tan\beta$ and charged Higgs mass $m_H$. }
\end{center}
\end{figure}

\newpage

\begin{figure}[ht]
\begin{center}
\epsfig{file=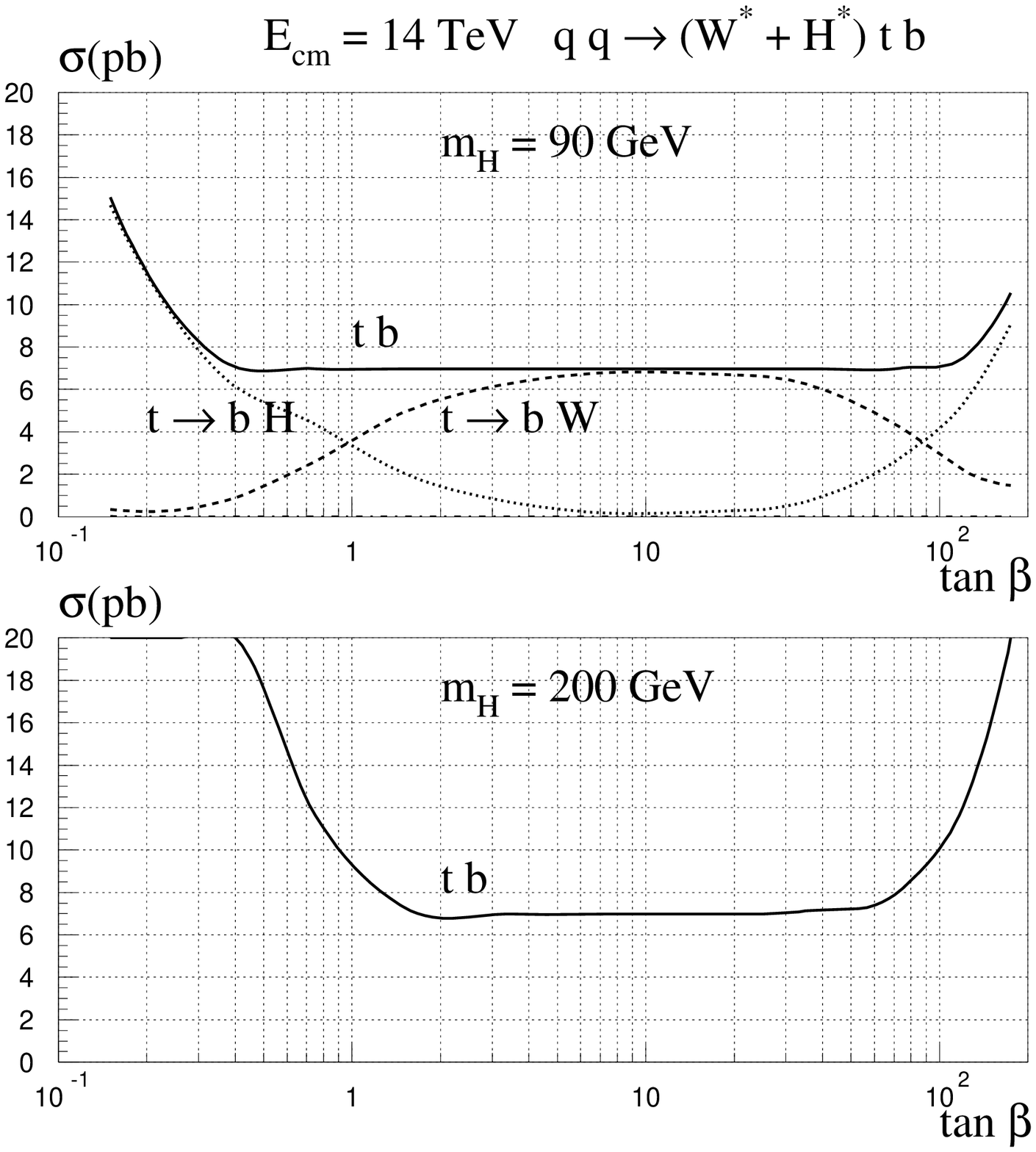,width=12cm,clip=}
\ccaption{}{The dependence of the ($t \bar b \, + \, \bar t b$)-pair 
production cross section 
versus $\tan\beta$ for two values of $m_H = 90$ and 200~GeV (the solid curves).
 The dashed and dotted curves correspond to the cross section production
times the branching ratio of $t$-quark decays into $b \wb$ and $b \cz$,
 respectively. }
\end{center}

\end{figure}
\begin{figure}[ht]
\begin{center}
\epsfig{file=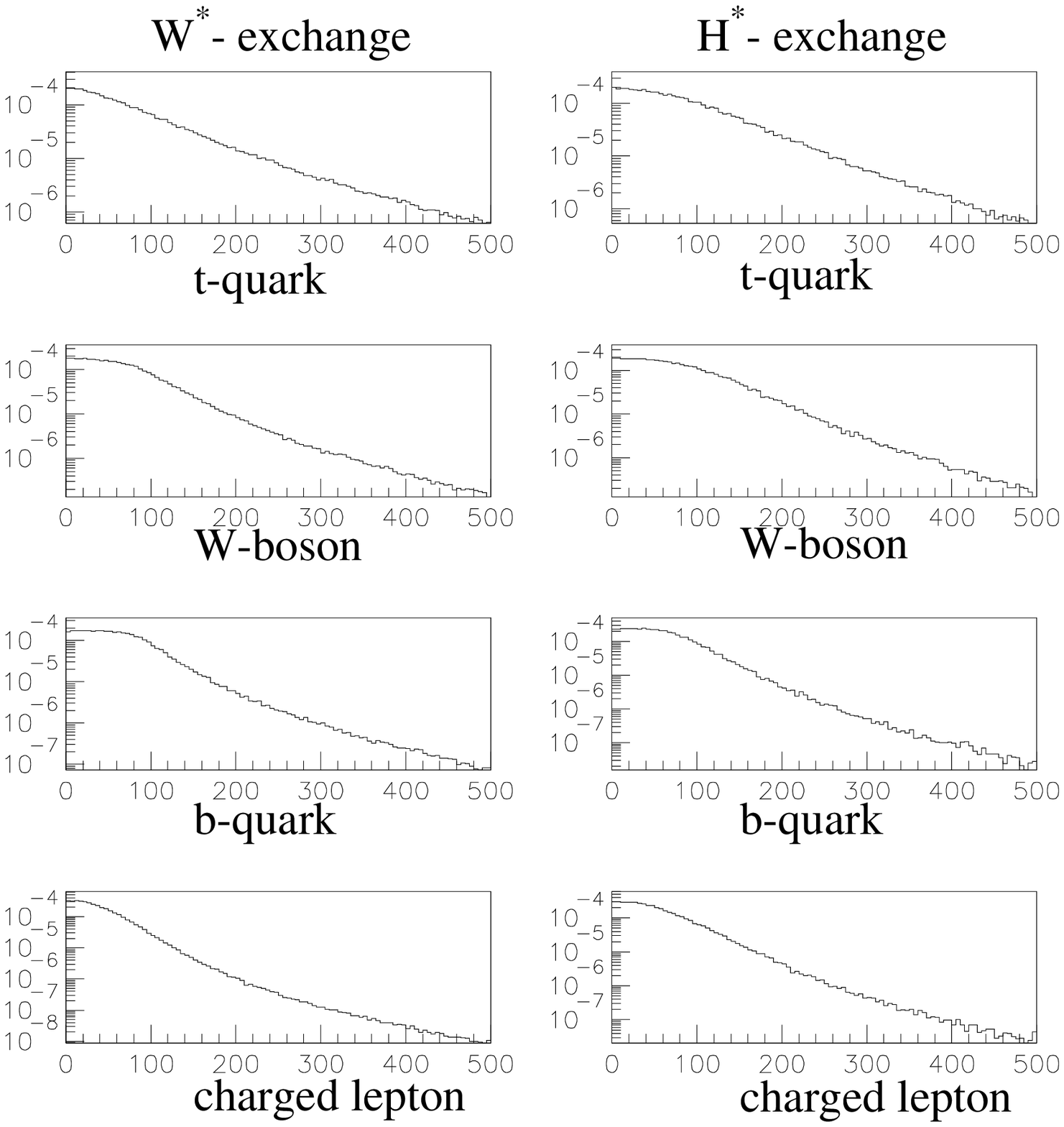,width=12cm,clip=}
\ccaption{}{The $\ptop$-distribution of the top quark and the products of its
decay $t \to b W (\to b l \nu)$ produced in reaction~(\ref{mainproc}).
The $\frac{d \sigma}{d \ptop}$ is in arbitrary units, while $\ptop$ (the $x$-axis) is
in~GeV. The symbol $W^*$ ($H^*$) corresponds to $t \bar b$-pair production
though $\wb$ ($\cz$)-exchange only. }
\end{center}
\end{figure}

\begin{figure}[ht]
\begin{center}
\epsfig{file=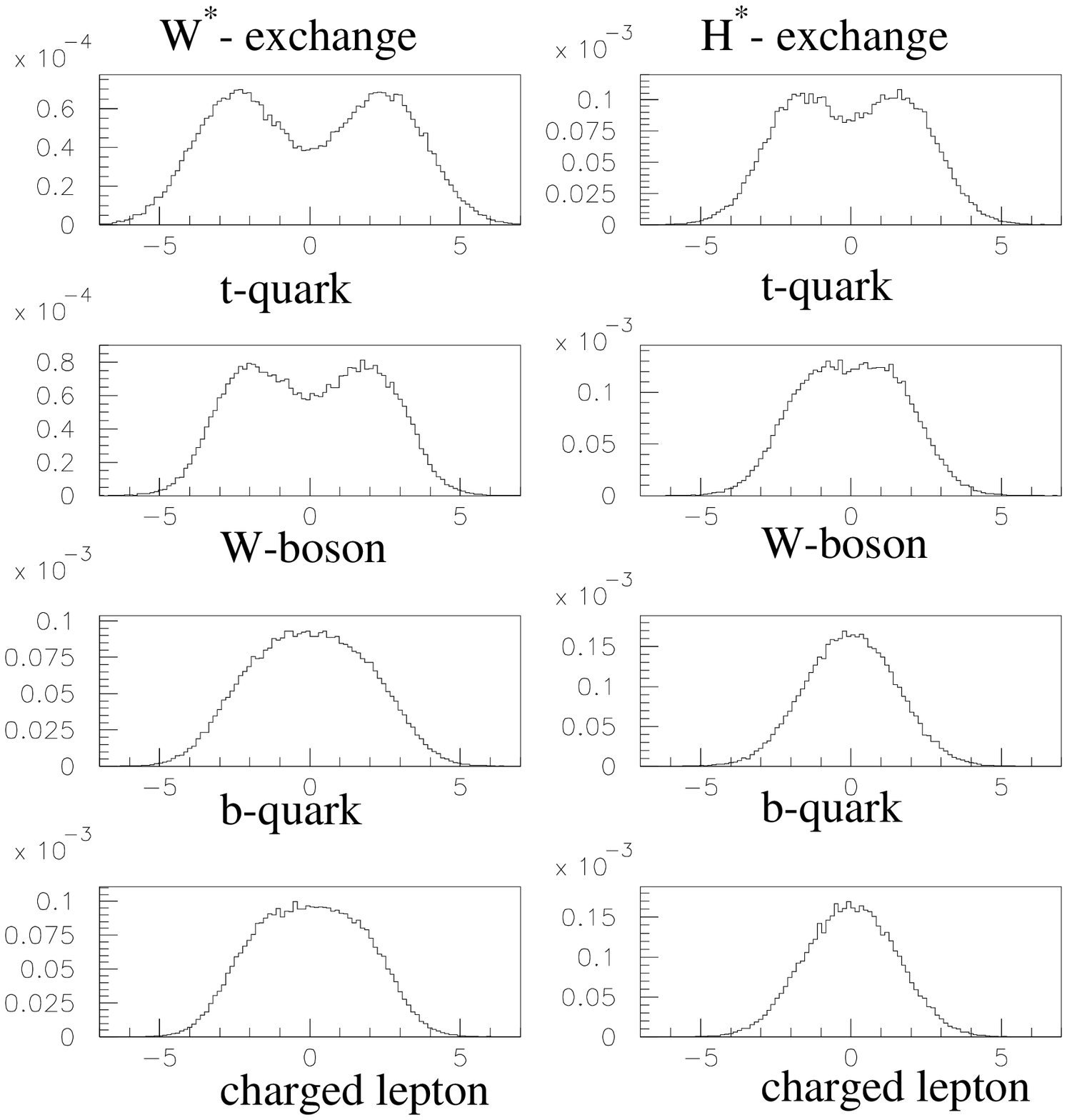,width=12cm,clip=}
\ccaption{}{The pseudorapidity, $\eta$, distributions of the $t$-quark
and its decay products. For additional explanation see Fig.4.}
\end{center}
\end{figure}

\begin{figure}[ht]
\begin{center}
\epsfig{file=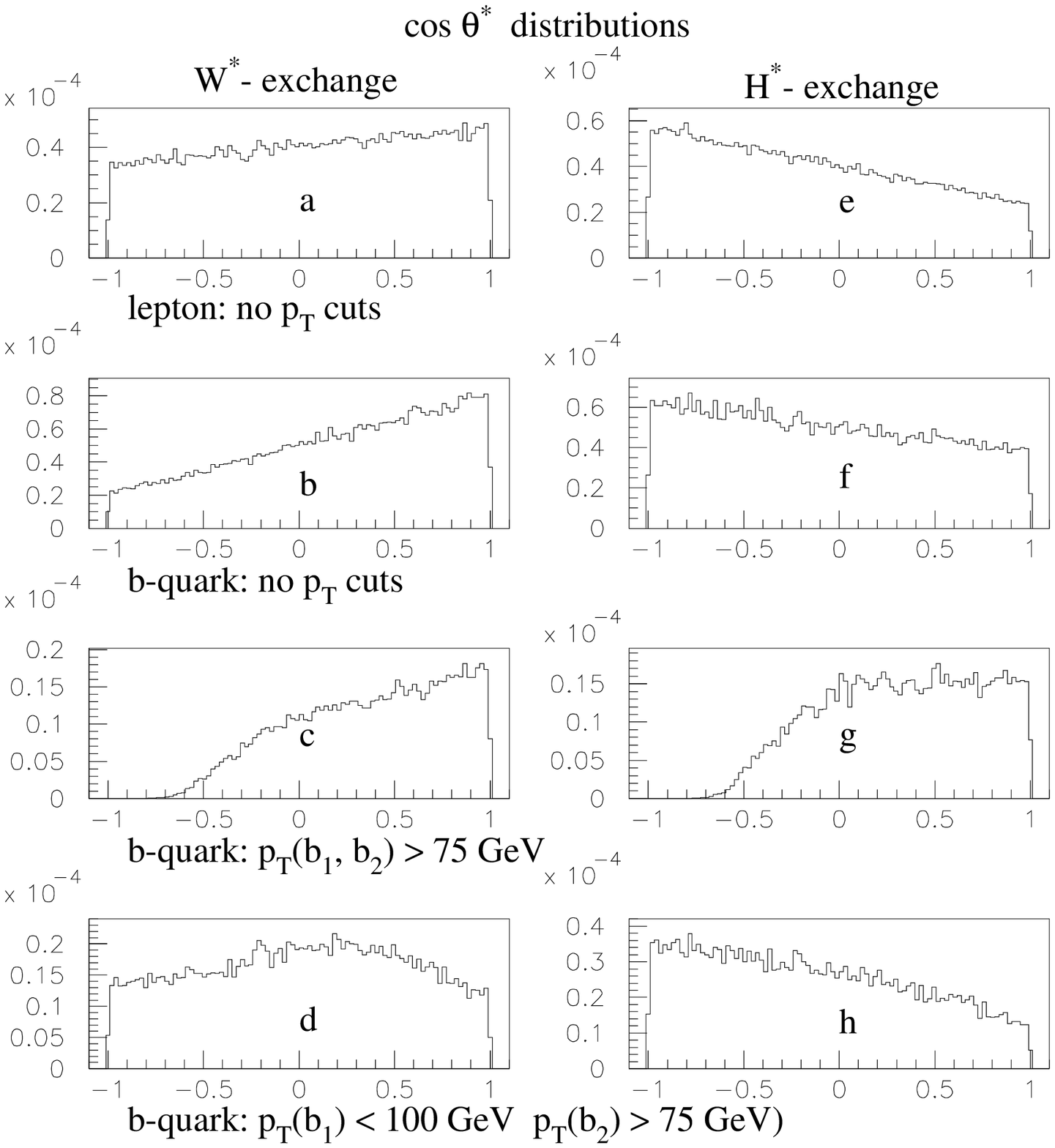,width=12cm,clip=}
\ccaption{}{The $\cos \vartheta^*$-distributions of the charged lepton and
$b$-quark originating from the $t$-quark decay. All the distributions are obtained
 from the evaluation of reaction~(\ref{mainproc}) at the parton level for
$\wb$-boson exchange ($W^*$) and $\cz$-boson contribution~($H^*$).
The histograms ``a, b, e, f'' are obtained without any cuts.
The H-cut~(\ref{hard}) and SH-cut~(\ref{newcut}) are used for
production of ``c, g'' and ``d,h'' histograms, respectively. }
\end{center}
\end{figure}

\begin{figure}[ht]
\begin{center}
\epsfig{file=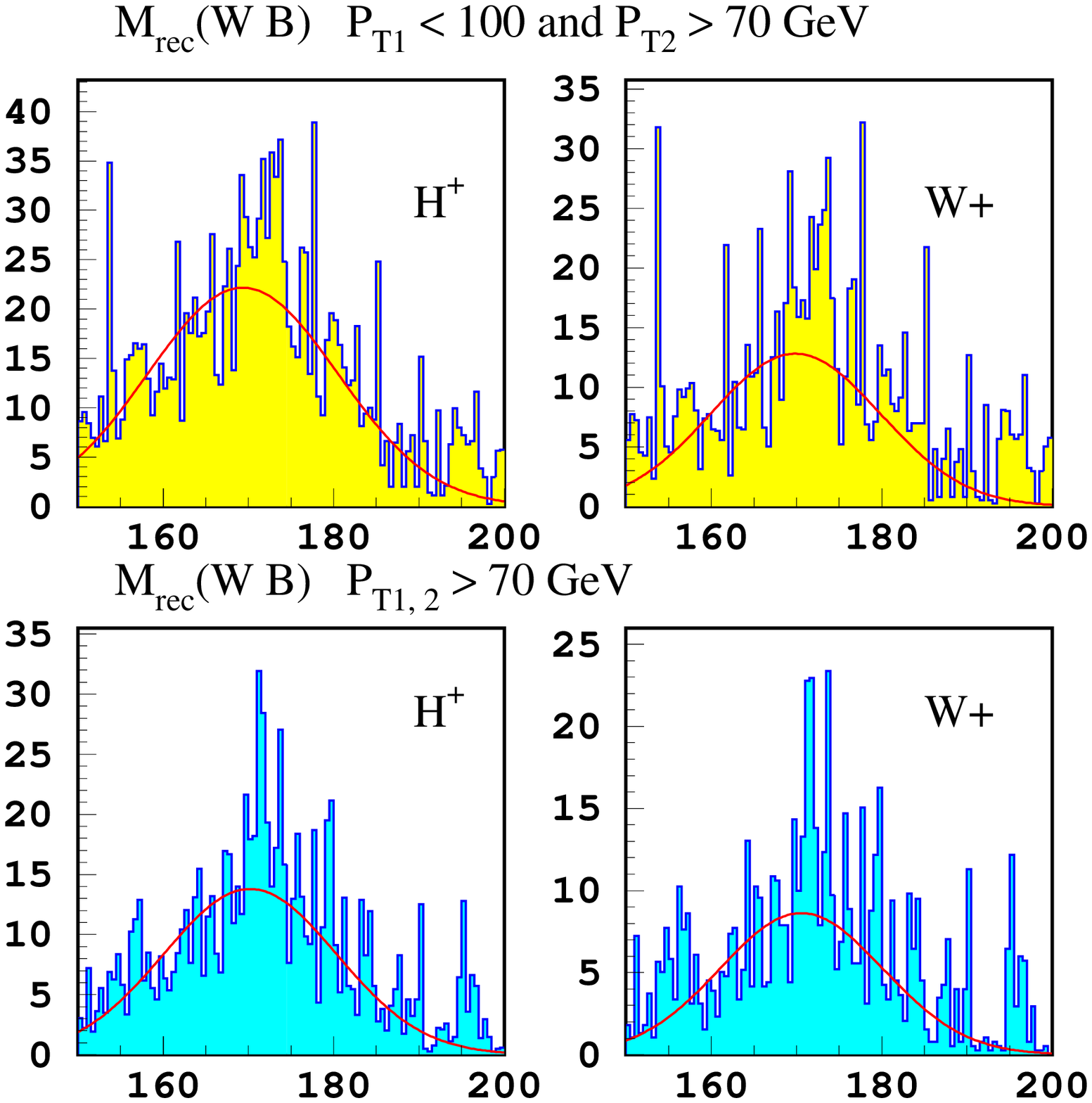,width=12cm,clip=}
\ccaption{}{The distributions (in the arbitrary units) on the invariant mass
of the ($W \, B$)-system, $M_{rec}$ (in GeV) resulted after application of all
 1--5 cuts. Two upper and lower histograms correspond to the application of
the new SH-cut~(\ref{newcut}) and old H-cut~(\ref{hard}),
respectively. The symbol $W^+$ corresponds to the SM
and background processes, while the symbol
$H^+$ refers to the $\cz$, the SM and background contributions. }
\end{center}
\end{figure}

\begin{figure}[ht]
\begin{center}
\epsfig{file=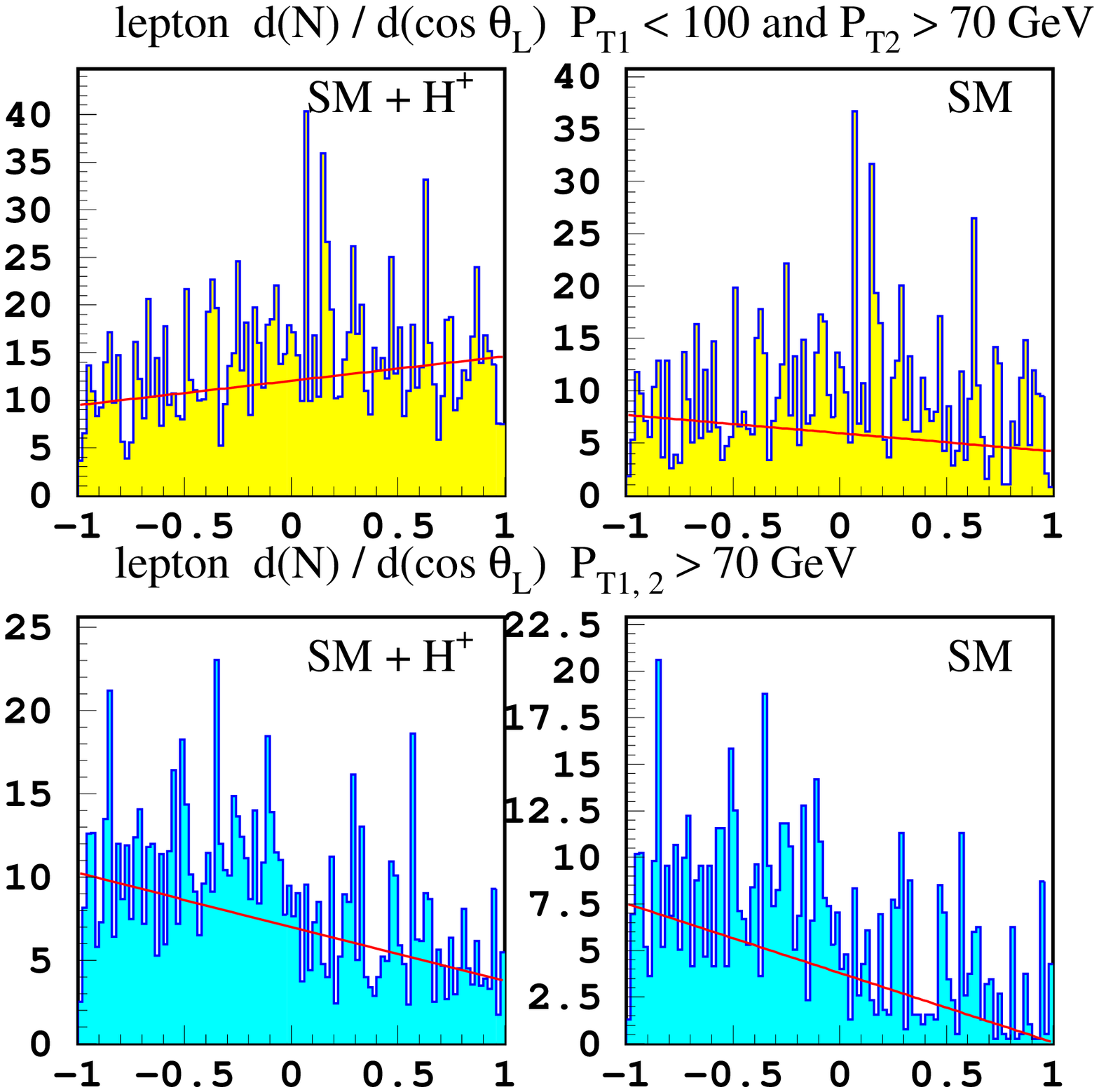,width=12cm,clip=}
\ccaption{}{The $\cos \vartheta^*$-distributions for the charged lepton.
The lines  correspond to the fit to linear dependence of~(\ref{fit}).  }
\end{center}
\end{figure}

\end{document}